# Three-body halos.
# II. From two- to three-body asymptotics

D.V. Fedorov[1], A.S. Jensen and K. Riisager

*Institute of Physics and Astronomy,
Aarhus University, DK-8000 Aarhus C*

**Abstract**

The large distance behavior of weakly bound three-body systems is investigated. The Schrödinger equation and the Faddeev equations are reformulated by an expansion in eigenfunctions of the angular part of a corresponding operator. The resulting coupled set of effective radial equations are then derived. Both two- and three-body asymptotic behavior are possible and their relative importance is studied for systems where subsystems may be bound. The system of two nucleons outside a core is studied numerically in detail and the character of possible halo structure is pointed out and investigated.

## 1  Introduction

The asymptotic behavior of halo nuclei [1, 2, 3], i.e. weakly bound and spatially extended nuclear systems, was recently discussed theoretically both for two-body systems [4] and for three-body systems where all subsystems are unbound [5, 6]. More complicated halo structures have also been anticipated [7, 8, 9]. These three-body systems have mean square radii which at the most diverge logarithmically with vanishing binding energy in contrast to two-body systems, where divergences are stronger and more abundant. This difference is rather peculiar, since a continous variation of the potential strengths clearly would lead from one type of asymptotics to the other. Thus a continous transition from two-body to three-body large-distance behavior must exist.

The interplay between the various asymptotics has to the best of our knowledge not been studied although the connection in principle is contained in the Faddeev equations [10, 11, 12]. Many, often very detailed, investigations of nuclear three-body systems have been published, see for example [13, 14, 15, 16, 17]. However, the proper mixture of two and three-body asymptotics is very difficult to include in practise for weakly bound systems. This insufficiency may be rather serious for halo nuclei where a mixture of two- and three-body behavior at large distance might be essential. Perhaps this plays a role in the astrophysically interesting $^8$B-nucleus [18]. This mixture is included in the treatment of the long range (attractive and repulsive) Coulomb interaction in loosely bound atomic systems [19, 20, 21]. However, the short-range nuclear interaction may behave differently. We shall employ the new procedure recently developed to study the Efimov effect in the coordinate space Faddeev equations [9].

The purpose of this paper is primarily to investigate the transition from two- to three-body asymptotic behavior for weakly bound three-body systems and secondly to advocate for more efficient methods [9] than previously used to solve the general three-body problem. As in our earlier papers [4, 5, 6], we want to establish the general features, qualitatively or quantitatively as best we can, and apply the results on realistic examples. We shall again for convenience assume that all three particles are inert and spinless. Thus all core degrees of freedom are frozen and all forces are assumed to be spin independent. Our model is therefore only an approximation of real systems in nature, but hopefully useful as a reference model, where the essential properties are exhibited.

The paper present is the second (after [6]) in a series of papers discussing various aspects of three-body halos. After the introduction we sketch in sections 2 and 3 practical theoretical descriptions based on the Schrödinger equation and the Faddeev equations, respectively. The proper mixture of two- and three-body asymptotic behavior for any system is then discussed in section 4. Illustrative numerical examples of this formulation are presented in section 5. Finally, we give the conclusions in section 6. Some of the mathematical definitions are collected in an appendix.

## 2  Schrödinger picture

The asymptotic large-distance behavior of three-body systems can be extracted from the general Faddeev equations [10, 11, 12]. However, the Schrödinger picture is much simpler and, as we shall see, also able to describe the three-body asymptotics correctly for configurations, where at most one binary subsystem is bound. We shall therefore in this section sketch a general method, based on the Schrödinger equation and well suited for universal applications. The method was used in atomic physics [19], but the asymptotics for short-range potentials were not investigated in detail. Recently also the triton was studied by use of the method [22].

---

[1] On leave from the Kurchatov Institute, 123182 Moscow, Russia

## 2.1 Radial equations

The Hamiltonian of the system, where the centre of mass kinetic energy is subtracted, is given by

$$H = \sum_{i=1}^{3} \frac{p_i^2}{2m_i} - \frac{P^2}{2M} + \sum_{i>j=1}^{3} V_{ij}(r_{ij}) , \qquad (1)$$

where $m_i$, $\mathbf{r}_i$ and $\mathbf{p}_i$ are mass, coordinate and momentum of the i'th particle, $V_{ij}$ are the two-body potentials, $P$ and $M$ are the total momentum and the total mass and $\mathbf{r}_{ij} = \mathbf{r}_i - \mathbf{r}_j$. The most convenient coordinates for the system are the Jacobi coordinates basically defined as the relative coordinates between two of the particles (x) and between their centre of mass and the third particle (y). The precise definitions are given in appendix A, where the corresponding three sets of hyperspherical coordinates ($\rho,\alpha,\Omega_x,\Omega_y$) are also defined. Here $\rho$ ($=\sqrt{x^2+y^2}$) is the generalized radial coordinate and $\alpha$, in the interval $[0,\pi/2]$, defines the relative size of x and y, $\Omega_x$ and $\Omega_y$ are the angles describing the directions of x and y. One of these sets of hyperspherical coordinates is in principle sufficient for a complete description. The volume element is given by $\rho^5 d\Omega d\rho$ ($d\Omega = \sin^2\alpha\cos^2\alpha d\alpha d\Omega_x d\Omega_y$) and the Hamiltonian in these variables is

$$H = T + \sum_{i>j=1}^{3} V_{ij} , \qquad (2)$$

where the kinetic energy operator $T$ is

$$T = T_\rho + \frac{\hbar^2}{2m} \frac{1}{\rho^2} \hat{\Lambda}^2, \; T_\rho = -\frac{\hbar^2}{2m} \left( \rho^{-5/2} \frac{\partial^2}{\partial \rho^2} \rho^{5/2} - \frac{1}{\rho^2} \frac{15}{4} \right) \qquad (3)$$

and $\hat{\Lambda}^2$ is the generalized angular momentum operator

$$\hat{\Lambda}^2 = -\frac{1}{\sin\alpha\cos\alpha} \frac{\partial^2}{\partial \alpha^2} \sin\alpha\cos\alpha + \frac{\hat{l}_x^2}{\sin^2\alpha} + \frac{\hat{l}_y^2}{\cos^2\alpha} - 4 . \qquad (4)$$

expressed in terms of the angular momentum operators $\hat{l}_x^2$ and $\hat{l}_y^2$ related to the x and y degrees of freedom.

The total wavefunction $\Psi$ of the three-body system (excluding the center of mass degrees of freedom) can now be expanded in terms of hyperangular functions

$$\Psi = \frac{1}{\rho^{5/2}} \sum_\lambda f_\lambda(\rho) \Phi_\lambda(\rho,\Omega) , \qquad (5)$$

where $\Phi_\lambda(\rho,\Omega)$ for each value of $\rho$ is a complete set as function of the variables $\Omega = \{\alpha,\Omega_x,\Omega_y\}$. These functions must be chosen to describe correctly the asymptotic behavior of the wavefunction and in particular the behavior corresponding to possible bound two-body subchannels.

These requirements are conveniently met by the eigenfunctions of the operator $\hat{\Lambda}^2$, the hyperspherical harmonics [25], in two cases, where the three-body binding energy is either large or small without bound subsystems. In the first case, the binding energy is large enough to confine the system to a size of the same order as the (short) ranges of the interactions [16]. In the second case, also with short-range potentials, the binding energy is small, but since each of the binary subsystems alone is not bound or nearly bound the hyperspherical harmonics provide the proper asymptotics [6].

If at the most one of the binary subsystems has bound or nearly bound states, the above requirements are met by the eigenfunctions of the operator $\hat{\lambda}$ defined by

$$\hat{\lambda} = \hat{\Lambda}^2 + \frac{2m\rho^2}{\hbar^2} \sum_{i>j=1}^{3} V_{ij} , \qquad (6)$$

where the eigenfunctions and eigenvalues $\lambda$ then are related by

$$\hat{\lambda}\Phi(\rho,\Omega) = \lambda(\rho)\Phi(\rho,\Omega) \ . \tag{7}$$

The expansion in eq. (5) is general and the particular choice in eq. (6) may be altered. If all potentials are omitted in the $\hat{\lambda}$-operator, the procedure is simply identical to the hypershercial expansion method [16]. The crucial point for our purpose is to include in $\hat{\lambda}$ the potential, which is able to bind the respective subsystem.

The Schrödinger equation which determines the total energy $E$ and the wavefunction $\Psi$ may now by use of eqs. (2)-(7) be rewritten as a coupled set of "radial" differential equations, i.e.

$$\left(-\frac{\partial^2}{\partial\rho^2} - \frac{2mE}{\hbar^2} + \frac{1}{\rho^2}\left(\lambda(\rho) + \frac{15}{4}\right)\right)f_\lambda + \sum_{\lambda'}\left(-2P_{\lambda\lambda'}\frac{\partial}{\partial\rho} - Q_{\lambda\lambda'}\right)f_{\lambda'} = 0, \tag{8}$$

where the functions $P$ and $Q$ are the following angular integrals:

$$P_{\lambda\lambda'}(\rho) \equiv \int d\Omega\, \Phi^*_\lambda(\rho,\Omega)\frac{\partial}{\partial\rho}\Phi_{\lambda'}(\rho,\Omega), \tag{9}$$

$$Q_{\lambda\lambda'}(\rho) \equiv \int d\Omega\, \Phi^*_\lambda(\rho,\Omega)\frac{\partial^2}{\partial\rho^2}\Phi_{\lambda'}(\rho,\Omega). \tag{10}$$

As mentioned already the procedure would not change, if some of the terms in $\hat{\lambda}$, for example one or two of the potentials, were left out and instead, after integration over angular variables, were included in the $\rho$-equation. The convergence rate and accuracy as such determines the optimal choice. However, we can already emphasize that the strongest of the potentials should be included in $\hat{\lambda}$, if we want to catch the corresponding two-body asymptotics. This is in contrast to the frequently applied expansion in hyperspherical harmonics, where all three potentials are left out in the $\hat{\lambda}$-operator [16, 25]. This results in either divergence or very slow convergence of the expansion in eq. (5), when the binding energy is small and one of the binary potentials is strong enough to hold a bound state.

## 2.2 Angular eigenvalue equation

The angular wavefunctions $\Phi_\lambda$ may now be expanded on the complete set of spherical harmonics for the directions of **x** and **y** in one of the sets of Jacobi coordinates. The expansion coefficients $\phi$ then depends on $\rho$ and $\alpha$, i.e.

$$\Phi_{\lambda LM}(\rho,\Omega) = \frac{1}{\sin\alpha\cos\alpha}\sum_{l_x l_y}\phi_{\lambda l_x l_y L}(\rho,\alpha)\mathbf{Y}^{LM}_{l_x l_y}(\Omega_x,\Omega_y) \ , \tag{11}$$

where $L$ and $M$ are the total angular momentum and its projection on the z-axis. The wavefunction $\mathbf{Y}$ is obtained by coupling $l_x$ and $l_y$:

$$\mathbf{Y}^{LM}_{l_x l_y}(\Omega_x,\Omega_y) \equiv [Y_{l_x}(\Omega_x)\cdot Y_{l_y}(\Omega_y)]_{LM} \ . \tag{12}$$

The coupled set of equations for $\phi_{\lambda l_x l_y L}$ is then

$$\left(-\frac{\partial^2}{\partial\alpha^2} + \frac{l_x(l_x+1)}{\sin^2\alpha} + \frac{l_y(l_y+1)}{\cos^2\alpha} - \lambda(\rho) - 4\right)\phi_{\lambda l_x l_y L}(\rho,\alpha) =$$
$$-\frac{2m\rho^2}{\hbar^2}\sum_{l'_x l'_y}W^{l'_x l'_y}_{l_x l_y L}(\rho,\alpha)\phi_{\lambda l'_x l'_y L}(\rho,\alpha) \ , \tag{13}$$

where the effective potentials $W$ are obtained by averaging the two-body potentials over all directions, i.e.

$$W^{l'_x l'_y}_{l_x l_y L}(\rho,\alpha) = \int d\Omega_x d\Omega_y\, \mathbf{Y}^{LM}_{l'_x l'_y}(\Omega_x,\Omega_y)\sum_{i>j=1}^{3}V_{ij}(\rho,\alpha,\Omega_x,\Omega_y)\mathbf{Y}^{LM}_{l_x l_y}(\Omega_x,\Omega_y) \ . \tag{14}$$

One of the potentials is in this coordinate system a function of **x**, whereas the other two potentials depend on both **x** and **y**. Therefore by using a finite number of angular momentum components in eq. (11) only one of the potentials can be accurately treated in eq. (13). A bound two-body state in one of the other potentials is then impossible to describe, since many angular momenta are required in the chosen set of coordinates.

## 2.3 Asymptotic behavior of the angular equation

The set of equations in eq. (13) obviously decouple at $\rho = 0$. Furthermore, in configurations with at most one bound two-body subsystem (corresponding to the x coordinate), the different angular momentum channels are also decoupled at $\rho = \infty$ assuming short-range central potentials. The asymptotic properties of systems, where two of the potentials are weak, can therefore be obtained by considering each set of angular momenta separately, i.e.

$$\left( -\frac{\partial^2}{\partial \alpha^2} + \frac{l_x(l_x+1)}{\sin^2 \alpha} + \frac{l_y(l_y+1)}{\cos^2 \alpha} + \frac{2m\rho^2}{\hbar^2} W^{l_x l_y}_{l_x l_y L}(\rho,\alpha) - 4 - \lambda(\rho) \right) \phi_{\lambda l_x l_y L}(\rho,\alpha) = 0 \; . \tag{15}$$

The expansion in eq. (11) implies that $\phi(\rho, \alpha = 0) = \phi(\rho, \alpha = \pi/2) = 0$ corresponding to infinitely high potential walls at these boundaries.

The characteristic properties of a short-range potential are as usual easiest illustrated by use of a square well potential. However, the results are correct for all short-range potentials in the limits of $\rho \to 0$ and at $\rho \to \infty$. We assume that eq. (15) includes only the strongest potential which is given by $V(x) = -S_0 \Theta(x < X_0)$, where $X_0 = a_x R_0$ is given in terms of the original square well radius $R_0$ and $a_x$ is related to the reduced masses of the system, see the appendix. The effective potential $W$ in the $\alpha$ variable is then also a square well potential easily seen to be given as $-\rho^2 \kappa_0^2 \Theta(\alpha < \alpha_0)$, where $\kappa_0^2 = 2mS_0/\hbar^2$ and $\alpha_0 = \arcsin(X_0/\rho)$ or $\pi/2$, when $X_0/\rho$ respectively is less than 1 or larger than 1.

We shall now consider the simplest case, $l_x = l_y = 0$, and study the quantized solutions for a given value of $\rho$. The wavefunction is given by

$$\phi(\rho,\alpha) \propto \begin{cases} \sin\left(\alpha\sqrt{\kappa_0^2 \rho^2 + \tilde{\lambda}}\right) & , \alpha < \alpha_0 \\ \sin\left((\alpha - \pi/2)\sqrt{\tilde{\lambda}}\right) & , \alpha > \alpha_0 \end{cases} \tag{16}$$

where the eigenvalue equation for $\tilde{\lambda} = \lambda + 4$ is found by matching at $\alpha_0$ which immediately gives

$$\sqrt{\kappa_0^2 \rho^2 + \tilde{\lambda}} \cot(\alpha_0 \sqrt{\kappa_0^2 \rho^2 + \tilde{\lambda}}) = \sqrt{\tilde{\lambda}} \cot\left((\alpha_0 - \pi/2)\sqrt{\tilde{\lambda}}\right) \; . \tag{17}$$

It is then easily seen that for $\rho < X_0$, where $\alpha_0 = \pi/2$, the eigenvalues are

$$\tilde{\lambda}_n(\rho) = 4n^2 - \kappa_0^2 \rho^2, \; n = 1, 2, 3, \ldots , \tag{18}$$

which in the limit of vanishing $\rho$ equals $4n^2$. Thus the spectrum of the eigenvalues, $\lambda_n = K(K+4)$, is the usual hyperspherical spectrum corresponding to the quantum numbers $K = 2n - 2$. The result in eq. (18) is valid for any short-range attractive potential when $\kappa_0^2$ is defined as the value obtained from the potential at the distance $x = 0$.

In the other limit of large $\rho$, we obtain a square well potential with decreasing radius ($\alpha_0 = X_0/\rho$), increasing depth ($S_0 \rho^2$) and entirely within the allowed interval between zero and $\pi/2$. The lowest positive eigenvalues, where $\lambda_n/\rho^2$ approaches zero, is by expansion to first order in $\alpha_0$ found from eq. (17) to be

$$\tilde{\lambda}_n(\rho) = 4n^2(1 - \frac{4a_x a}{\pi \rho}), \; n = 1, 2, 3, \ldots , \tag{19}$$

which holds for any short-range potential, when $a$ is the scattering length, which for the square well potential is given by

$$a = R_0 \left( \frac{\tan(\kappa_0 X_0)}{\kappa_0 X_0} - 1 \right) \; . \tag{20}$$

As in the limit of vanishing $\rho$, the eigenvalue $\tilde{\lambda}$ converges again towards $4n^2$ for $\rho \to \infty$. Only one of the three scattering lengths enter eq. (19) which again shows that the Schrödinger picture is unable to describe all three two-body subsystems on an equal basis.

When the scattering length is infinitely large a singularity appears at large $\rho$. The eigenvalues $\tilde{\lambda}$ converge in this case towards $(2n-1)^2$ for $\rho \to \infty$ as seen from eq. (17).

It is now straightforward to extend this discussion to finite values of $l_x$ and $l_y$, which introduces a repulsive potential in the $\alpha$ degree of freedom. The effective Hamiltonian is then changed by the two additional angular momentum terms. The general solutions are hypergeometric functions both below and above $\alpha_0$. Matching provide the eigenvalue equation analogous to eq. (17). As for $l_x = l_y = 0$, the positive energy solutions again have the hyperharmonical spectrum for both small and large $\rho$, but now with $K = 2n - 2 + l_x + l_y$.

Negative eigenvalues of $\hat{\lambda}$ also sometimes occur at large distances $\rho$, where $\alpha_0$ is small, $\cos\alpha \approx 1$ and $\sin\alpha \approx \alpha$ inside the well. The differential equation for such solutions may by use of eq. (15) be rewritten

$$\left(-\frac{\partial^2}{\partial z^2} + \frac{l_x(l_x+1)}{z^2} + \frac{2m}{\hbar^2}W(z/a_x)\right)\phi(z) = \frac{2mE_x}{\hbar^2}\phi(z) \ , \tag{21}$$

where the variables are changed to $z = \rho\alpha$ and $\tilde{\lambda} - l_y(l_y + 1) \equiv 2m\rho^2 E_x/\hbar^2$ and where we assumed that $\rho$ is large compared to the characteristic radius $R_0$, where the potential has vanished. Eq. (21) is valid to an accuracy of $R_0/\rho$, since $W$ vanishes for large $\alpha$ and $z \approx x$ for small $\alpha$. The boundary condition $\phi(z = 0) = \phi(z = \rho\pi/2 \to \infty) = 0$ then completes the equivalence to the corresponding problem for two interacting particles, where the negative eigenvalues $\tilde{\lambda}$ now are related to the two-body bound state energy $E_x$.

The effective potential in $\alpha$-space has, in the limit of large $\rho$, exactly the same number of bound states (perhaps none) as the original two-body potential. These solutions lead to finite negative asymptotic values of $\tilde{\lambda}/\rho^2$ and the bound state wavefunctions are similar to the corresponding two-body solutions, oscillating inside and decaying outside the square well. The actual wavefunction is for $l_x = l_y = 0$ obtained from eq. (16) by allowing a negative value of $\tilde{\lambda}$, which changes the sine-function in the outer region into a hyperbolic sine-function.

Inclusion of the previously neglected two-body potentials modify the effective potential in $\hat{\lambda}$. For small values of $\rho$, where all distances also are small and $\alpha_0 = \pi/2$, the individual potential strengths simply should be added and the eigenvalues are again given by eq. (18) with $\kappa_0^2$ substituted by the sum of all three contributions. In the other limit of large $\rho$, where the potentials in $\alpha$-space become narrow and deep (width $\propto 1/\rho$, depth $\propto \rho^2$), eigenvalues and wavefunctions remain unchanged. It is not possible in the Schrödinger picture to describe asymptotics related to more than one binary bound subsystem. The reason is that components in the wavefunction corresponding to both bound binary states are necessary. The Schrödinger equation is formulated by use of one set of coordinates and the description of one bound two-body state expressed in terms of another ("not natural") set of Jacobi coordinates requires a large number of angular momentum components. This number increases linearly with $\rho$ and as a consequence infinitely many angular momenta would be needed for $\rho \to \infty$.

The analytical results obtained here for the square well are typical for any short-range potential. All arguments are identical, if "outside the square well" is replaced by "outside the range of the potential", i.e. at a distance much larger than the characteristic fall-off length of the potential. Thus we conclude, that a three-body system interacting via short-range potentials and with at most one subsystem strong enough to hold bound states, the asymptotic behavior of the angular spectrum is hyperharmonical at both small and large $\rho$. In addition there is at large $\rho$, for every bound binary subsystem, a parabolically diverging angular eigenvalue $2mE_x\rho^2/\hbar^2$. When the scattering length is infinitely large the hyperharmonical spectrum at large $\rho$ is shifted downwards by one unit in the $K$ quantum number.

A long-range repulsive two-body potential proportional to $r^{-\nu}$ ($\nu < 2$) contributes basically to the effective angular potential by a term proportional to $\rho^{2-\nu}$, see eq. (13). The contribution to the angular eigenvalue $\tilde{\lambda}$ vanishes then at $\rho = 0$ and the hyperspherical spectrum remains. At large distances we must again distinguish between positive and negative values of $\tilde{\lambda}$. As for short-range potentials we arrive for bound states also now at the asymptotic equation in eq. (21), where the effective potential $W$ includes the contribution from the long-range potential. Thus the parabolic divergence towards $-\infty$ and the bound state energy remains related in the same way. The positive eigenvalues are increasing as $\rho^{2-\nu}$ in the asymptotic region as seen from eq. (15), where the strength of the effective potential now is proportional to this quantity.

## 3  Faddeev picture

There is an intriguing variety of possible structures of a quantum mechanical three-body system. Very peculiar configurations may arise especially at small binding energies, where large spatial distances of the system is allowed. For example, if two of the subsystems can form weakly bound states, one configuration in the three-body system is a coherent superposition of two different (mutually exclusive) configurations, where two particles are close and the third far away. An efficient description of such states demands at least two components in the total wavefunction, one for each of the asymptotic two-body configurations. In general three components are needed. Each of the components can be rather simple, while the total wavefunction has a very complicated structure due to the coherent interplay of different components. These subtleties are better described in the Faddeev equations [10, 11, 12] and we shall in this section sketch a general method, which is well suited for universal applications as well as investigations of asymptotic properties. It has already been applied on systems with all angular momenta equal to zero [9].

## 3.1 Radial equations

The total wavefunction $\Psi$ of the three-body system is therefore written as a sum of three components each expressed in terms of one of the three different sets of Jacobi coordinates:

$$\Psi = \sum_{i=1}^{3} \psi^{(i)}(\mathbf{x}_i, \mathbf{y}_i). \tag{22}$$

This three-component wavefunction is flexible and allows a description of different three-body structures by means of rather few angular momenta in each component. These wavefunctions satisfy the three Faddeev equations [10]

$$(T - E)\psi^{(i)} + V_{jk}(\psi^{(i)} + \psi^{(j)} + \psi^{(k)}) = 0, \tag{23}$$

where $E$ is the total energy, $T$ is the kinetic energy operator and $\{i, j, k\}$ is a cyclic permutation of $\{1, 2, 3\}$.

All solutions of the Faddeev equations satisfy, via eq. (22), also the Schrödinger equation. Clearly $\Psi$ can vanish identically even when the three components $\psi^{(i)}$ individually are non-zero functions. In fact, these additional solutions are the trivial solutions of the corresponding Schrödinger equation. The Faddeev equations and the Schrödinger equation provide the same solutions, if only the total solution $\Psi = \psi^1 + \psi^2 + \psi^3$ is considered.

Each component is now for each $\rho$ expanded like in eq. (5) on a complete set of generalized angular functions $\Phi_\lambda^{(i)}(\rho, \Omega_i)$. The radial expansion coefficients $f_\lambda(\rho)$ are component independent, since $\rho$ is invariant for all three Jacobi sets. If furthermore the angular functions are chosen as the eigenfunctions of the angular part of the Faddeev equations, we arrive again at the coupled set of radial equations in eq. (8). The only difference is that $\Phi$ in the scalar products of eqs. (9) and (10) now should be substituted by the sum of the three components, i.e. $\Phi = \Phi^1 + \Phi^2 + \Phi^3$.

## 3.2 Angular eigenvalue equation

The wavefunctions $\Phi_\lambda^{(i)}$ can be expanded on the complete set of spherical harmonics for the directions of $\mathbf{x}$ and $\mathbf{y}$. The expansion coefficients $\phi$ then depends on $\rho$ and $\alpha_i$, i.e.

$$\Phi_{\lambda LM}^{(i)}(\rho, \Omega_i) = \frac{1}{\sin \alpha_i \cos \alpha_i} \sum_{l_{ix} l_{iy}} \phi_{\lambda l_{ix} l_{iy} L}^{(i)}(\rho, \alpha_i) \mathbf{Y}_{l_{ix} l_{iy}}^{LM}(\Omega_{ix}, \Omega_{iy}), \tag{24}$$

where $\mathbf{Y}$ is defined in eq. (12) and $L$ and $M$ are the total angular momentum and its projection on the z-axis. The coupled set of equations for $\phi_{\lambda l_{ix} l_{iy} L}^{(i)}$ is found to be

$$\left(-\frac{\partial^2}{\partial \alpha_i^2} + \frac{l_{ix}(l_{ix}+1)}{\sin^2 \alpha_i} + \frac{l_{iy}(l_{iy}+1)}{\cos^2 \alpha_i} - 4 - \lambda(\rho)\right) \phi_{\lambda l_{ix} l_{iy} L}^{(i)}(\rho, \alpha_i) =$$
$$-\frac{2m\rho^2}{\hbar^2} \sum_{j=1}^{3} \sum_{l_{jx} l_{jy}} W_{i l_{ix} l_{iy} L}^{l_{jx} l_{jy}}(\rho, \alpha_i) \phi_{\lambda l_{jx} l_{jy} L}^{(j \to i)}(\rho, \alpha_i) \tag{25}$$

where $\phi^{(j \to i)}$ is defined by

$$\sum_{l_x l_y} \phi_{\lambda l_x l_y L}^{(j)}(\rho, \alpha_j) \mathbf{Y}_{l_x l_y}^{LM}(\Omega_{jx}, \Omega_{jy}) \equiv \sum_{l_x l_y} \phi_{\lambda l_x l_y L}^{(j \to i)}(\rho, \alpha_i) \mathbf{Y}_{l_x l_y}^{LM}(\Omega_{ix}, \Omega_{iy}) \tag{26}$$

This transformation "rotates" from the j'th to the i'th Jacobi coordinate system.

The effective potentials $W_i$ are obtained by averaging the two-body potentials over all directions, i.e.

$$W_{i l_x l_y L}^{l'_x l'_y}(\rho, \alpha_i) = \int d\Omega_x d\Omega_y \, \mathbf{Y}_{l'_x l'_y}^{LM*}(\Omega_x, \Omega_y) V_i(\rho, \alpha_i, \Omega_x, \Omega_y) \mathbf{Y}_{l_x l_y}^{LM}(\Omega_x, \Omega_y), \tag{27}$$

where $V_i$ is the potential acting between the two particles different from that labeled i.

## 3.3 Asymptotic behavior of the angular equation

The quantized solutions of eq. (25) clearly decouple at $\rho = 0$. For short-range potentials they also decouple at $\rho = \infty$ when $l_{ix} > 0$ (see section 2.3) and the asymptotic properties of such systems can for each component therefore be obtained from eq. (15). The remaining coupled equations, where $l_{ix} = 0$ in all three subsystems, are very essential as they provide the dominating part of the large distance behavior. However, eigenvalues and eigenfunctions are rather simple in both limits of large and small $\rho$.

For small $\rho$, square well potentials and $l_y = 0$, we obtain again for each $\phi^{(i)}$ the solutions in eq. (16), where $\kappa_0^2$ now is the sum found by adding the three potentials at zero distance. Clearly the form in eqs. (17) and (18) then follow immediately. At large $\rho$ the solution for square wells and all $l_{ix} = 0$ is instead of the form

$$\phi^{(i)}(\rho, \alpha_i) \propto \begin{cases} b_i \sin\left(\alpha_i \sqrt{\kappa_i^2 \rho^2 + \tilde{\lambda}}\right) + c_i \sin\left(\alpha_i \sqrt{\tilde{\lambda}}\right) & , \alpha_i < \alpha_{i0} \\ a_i \sin\left((\alpha_i - \pi/2)\sqrt{\tilde{\lambda}}\right) & , \alpha_i > \alpha_{i0} \end{cases} \quad (28)$$

where $\kappa_i^2 = 2mS_i/\hbar^2$, $\alpha_{i0} = \arcsin(X_i/\rho)$ and the nine coefficients $a_i$, $c_i$ and $b_i$ are related by six linear constraints obtained from eq. (25). Matching conditions at $\alpha_{i0}$ then provide positive eigenvalues, which asymptotically behave like eq. (19), where the scattering length $a$ times $a_x$ now must be replaced by the sum of the three individual quantities [9].

Again negative eigenvalues $\tilde{\lambda}$ at large distances $\rho$ also sometimes arise from the matching conditions as the signature of either bound binary subsystems or two or three infinitely large scattering lengths. The effective potential in $\alpha$-space still has exactly the same number of bound states (perhaps none) as all the original two-body potentials together. These solutions are decoupled at $\rho = \infty$ and lead to finite negative asymptotic values of $\tilde{\lambda}/\rho^2 = 2mE_x/\hbar^2$ corresponding to the energy $E_x$ of the bound subsystem.

A long-range repulsive two-body potential proportional to $r^{-\nu}$ ($\nu < 2$) changes the effective potential at large distances. The strength as well as the positive eigenvalues increase proportional to $\rho^{2-\nu}$ whereas the negative eigenvalues diverge parabolically.

## 4 Asymptotic large-distance behavior

The asymptotic properties of the wavefunctions can now be extracted by use of the information in the previous subsections. The radial set of equations in eq. (8) is valid both in the Schrödinger and the Faddeev picture. They are decoupled in the limit of large $\rho$ as seen from the fact that the non-diagonal parts of both P and Q approaches zero faster than $\rho^{-3}$. The decisive quantity $\lambda(\rho)$ is obtained by solving eqs. (13) or (25), which has the hyperspherical eigenvalue spectrum at $\rho = 0$, i.e. $\lambda = K(K + 4)$ where $K$ is a non-negative integer. The spectrum at large $\rho$ depends on the number of bound states in the binary subsystems.

The sequential structure of our procedure suggests definitions of partial energies, $E_x(\rho) = \hbar^2 \tilde{\lambda}/(2m\rho^2)$ and $E_y(\rho) = E - E_x$, related to the "angular" and radial degrees of freedom. They are both functions of $\rho$, but added up they must result in the total energy $E$, which in our considerations almost exclusively is assumed to be negative corresponding to a three-body bound state. The decay constants connected with the energies are given by $\kappa_x^2 = -2mE_x/\hbar^2$, $\kappa_y^2 = -2mE_y/\hbar^2$ and $\kappa^2 = -2mE/\hbar^2$.

A schematic overview of the various possibilities for a three-body system interacting through two-body potentials of varying strengths is shown in fig. 1. Instead of showing a three dimensional contour diagram, we projected on the two different planes defined by one vanishing potential and two identical potentials. In region I, where all potentials are weak, all energies, total as well as partial, are positive for all $\rho$ and the system and its subsystems are all unbound. This is not an interesting region in the present context. The other regions will be discussed in the following subsections.

### 4.1 No bound binary subsystems

In region II we have a bound three-body state of energy $E(< 0)$, whereas all the two-body potentials are too weak to support bound states. The $\lambda$ spectrum at infinity is identical to that of $\rho = 0$, i.e. $\lambda_n(\rho = \infty) = \lambda_n(\rho = 0) = K(K + 4)$. The partial energies $E_x$ and $E_y$ for the lowest eigenvalue with the quantum number in question are respectively negative and positive in at least an intermediate region of $\rho$ values. The asymptotic form of the related (diagonal) radial equation, see eq. (8), is then

$$\left(-\frac{\partial^2}{\partial \rho^2} + \kappa^2 + \frac{(K + 3/2)(K + 5/2)}{\rho^2}\right) f_\lambda = 0 \ . \quad (29)$$

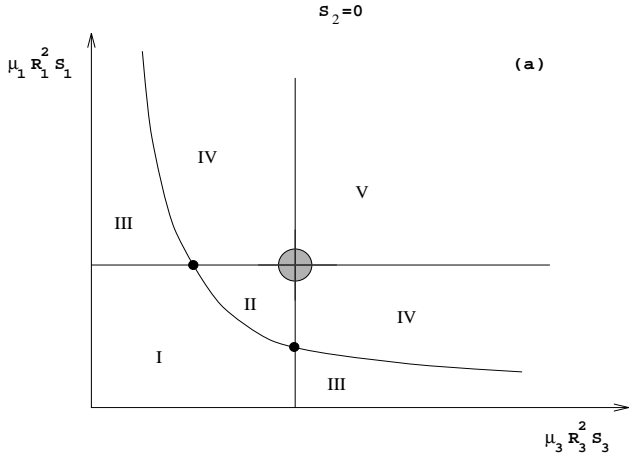
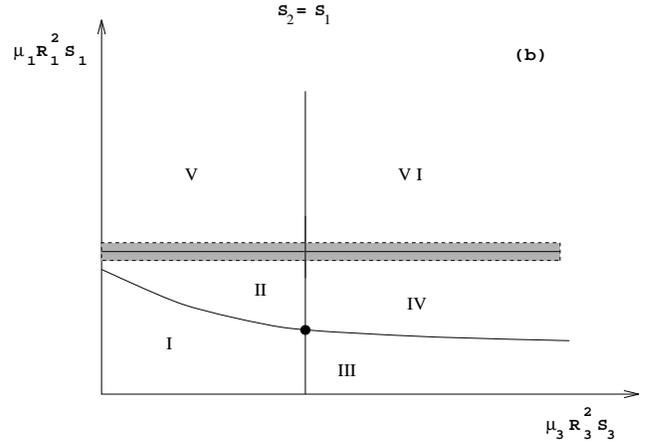

Figure 1a. Schematic contour diagram of the bound three-body state as function of the decisive depth and radius combinations. One of the potentials is assumed to be identically zero. The depths, radii and reduced masses are denoted, respectively $S$, $R$ and $\mu$ with indices for the different coordinates. The full curve is dividing the regions, where the three-body system is either bound or unbound. In region I the system as well as all subsystems is unbound. In regions III and IV one subsystem is bound, whereas two subsystems are bound in region V. The total three-body system is also bound in regions II and IV. In region III one subsystem is bound and the third particle unbound. The dark circles are the special points, where the asymptotics is determined by the direction of approach. The shaded circle indicates the singular Efimov point.

Figure 1b. The same as figure 1a when two of the potentials are identical. All three subsystems are bound in region VI and the Efimov singularity is now indicated by the horizontal shaded rod.

We are left with the same asymptotic structure as found by expansion in hyperspherical harmonics.

The asymptotic behavior for vanishing binding energy (towards region I in fig.1) of the radial moments, $<\rho^n>$, $<x^n>$ and $<y^n>$, are derived and discussed in [6]. We shall not repeat this derivation here, but only quote a few pertinent results. The asymptotic potential has the form of a centrifugal barrier term with angular momentum quantum number $l^* = K + 3/2$. The behavior of the moments are therefore determined by the relative size of $n$ and $2l^* - 1$, see [4, 5]. The moments converge or diverge respectively for $n < 2K + 2 (= 2l^* - 1)$ or $n \geq 2K + 2$. Thus the first ($n = 0, 1$) moments of unnormalized wavefunctions are always convergent, which means that the probability distribution is confined to finite distances. Divergences of the second moments only occur when all angular momenta are zero and then the divergences are logarithmic. This is the genuine three-body asymptotic structure.

The eigenvalues $\tilde{\lambda}$ are smaller than the hyperspherical result (4 for zero angular momenta) for intermediate values of $\rho$, see eqs. (18) and (19). The closer we are to a bound subsystem (towards the regions IV or V in fig.1), the smaller is the minimum value of $\tilde{\lambda}$. In the equivalent hyperspherical harmonical expansion this results in a much slower convergence and a need for many terms, because the asymptotic region then only is reached at very small binding energies. The three-body state already begins to "feel" the bound state and consequently diverge stronger with energy (as two-body systems) before reaching the (slower) extreme three-body asymptotic region.

When a bound state of zero energy eventually is reached and the scattering length is infinitely large, the asymptotic form of the radial equation still remains as eq. (29), but with $K$ shifted downwards by 1. Thus the lowest value, $K = -1$ corresponds to an effective angular momentum $l^*$ of $1/2$, which means that the wavefunction itself diverges logarithmically and the second moment as $1/(-E\ln(-E))$.

A long-range repulsive two-body potential proportional to $r^{-\nu}$ ($\nu < 2$) changes the angular eigenvalues at large distances into functions increasing like $\rho^{2-\nu}$. This in turn confine the wavefunctions and lead to finite mean square radii even for vanishing binding energy.

## 4.2 One bound binary subsystem

In the regions III and IV, where only the strongest potential is able to hold a bound state, we have for the lowest solution that $E < 0$, $\tilde{\lambda}(\rho \to \infty) < 0$ and $E_x(\rho \to \infty) < 0$. As seen from eqs. (16) and (21), the angular wavefunction in eq. (15) is exponentially decaying, i.e. $\phi(\rho, \alpha) \propto \exp(-\alpha\rho\kappa_x)$. This is for small $\alpha$ the exponential tail, $\exp(-x\kappa_x)$, of a two-body bound state for a system with coordinate $x$, energy $E_x$ and reduced mass $m$ or equivalently a system with coordinate $x/a_x$, energy $E_x$ and reduced mass $ma_x{}^2$.

When one subsystem have bound states of energies $E_x$, the lowest $\tilde{\lambda}$ values diverge as $\tilde{\lambda}(\rho \to \infty) \to -|E_x|\rho^2 2m/\hbar^2$, where the possibly different $E_x$ values each correspond to one eigenvalue. The coordinates $x$ and $y$ now refer to the set of Jacobi coordinates in which $x$ is the relative coordinate of the bound subsystem. The higher lying $\tilde{\lambda}$ values again converge to the hyperspherical spectrum for $\rho \to \infty$. The asymptotic form of the related (diagonal) radial equation, see eq. (8), is then

$$\left(-\frac{\partial^2}{\partial \rho^2} + \kappa^2 - \kappa_x^2 + \frac{l_y(l_y+1)}{\rho^2}\right) f_\lambda = 0 , \tag{30}$$

where the "centrifugal barrier height", $l_y(l_y + 1)$, may be found by a perturbation treatment, which in both eq. (8) and eq. (21) must include terms up to the order $\rho^{-2}$. Eq. (30) is the asymptotic form of a radial two-body equation describing the relative motion of a particle of angular momentum $l_y$ and "reduced" energy $\kappa^2 - \kappa_x^2$.

When $E_y(\rho \to \infty) = E - E_x$ is positive (region III), the lowest solution of eq. (30) is unbound, which means that the ground state of the three-body system is a bound two-body state in the $x$ degree of freedom, while the third particle recides at infinity or takes up the remaining positive energy in a continuum state. This is a rather uninteresting solution in the present context.

In region IV, where $E_y = E - E_x$ is negative for large values of $\rho$, we obtain immediately from eq. (30) that the bound state $f$ is exponentially decaying at large distance, i.e. $f \propto \exp(-\rho\kappa_y)$. The total wavefunction $\Phi$ at large $\rho$ and small $\alpha$, where $\rho \approx y$, is then behaving like $f\phi \propto \exp(-x\kappa_x - y\kappa_y)$. This is the asymptotic form of a three-body wavefunction, when a two-body subsystem is bound, see ref. [11]. It describes the tail of two asymptotically independent bound states: one for the $x$ and one for the $y$ degree of freedom. (Use of the correct reduced masses with the related coordinate changes leave this expression unchanged.) In fact this asymptotic behavior is correct for large $\rho$ and arbitrary $\alpha$, since large $\alpha$ implies large $x$ and consequently vanishing $f\phi$.

The small binding limit in region IV (towards region III in fig. 1) now means that the total three-body energy $E$ approaches the energy $E_x$ of the bound two-body system. The radial moments of interest in this case are $<x^n>$ and $<y^n>$, which as seen from eq. (30) and the asymptotics described above, independently behave like in a two-body system [4]. Thus $<x^n>$ approaches its value for the bound state while $<y^n>$ converges for $n < 2l_y - 1$ and diverges for $n \geq 2l_y - 1$, i.e. for example $<y^2> \approx <\rho^2> \to 1/E_y$ for $l_y = 0$.

In particular, let us focus the attention on the singular points in fig. 1, where the regions I,II,III and IV are touching each other. The asymptotic behavior of the wavefunction depends on the region from which the point is approached, i.e. either two- or three-body asymptotics. On the other hand when moving from region II to region IV all quantities are smooth. When one singular point is approached exactly on the dividing line between II and IV, where the scattering length is infinitely large, the radial equation has to be modified as described in subsection 4.1.

It should be emphasized in general that $E_x$ always is positive at smaller values of $\rho$, even when a two-body bound state is possible. This in turn influences the behavior at smaller $\rho$ of the solution $f$ to the differential equation, which once more then resembles the equation arising after the expansion in hyperspherical harmonics. Thus, when the binary subsystem is bound, we find two-body asymptotics at large $\rho$, whereas the usual three-body behavior smoothly is obtained by decreasing $\rho$. The method using the Schrödinger picture is apparently able to describe the proper mixture of these two types of large distance behavior. On the other hand it is also essential to realize that, even if all three two-body potentials were included in the operator $\hat{\lambda}$, only the possible two-body asymptotics in one of the subsystems can be picked up by solving eq. (13).

A long-range repulsive two-body potential proportional to $r^{-\nu}$ ($\nu < 2$) changes the angular eigenvalues at large distances into functions increasing like $\rho^{2-\nu}$, except for the lowest lying values (where the $\rho^{2-\nu}$ contribution is overrun by the main term proportional to $\rho^2$) corresponding to the total number of bound states in all the binary subsystems. The total wavefunctions behave accordingly, i.e. they are spatially confined for all the upsloping values, all radial moments are finite, and their asymptotics remain unchanged for the downsloping values.

## 4.3 Two or three bound binary subsystems

In region V and VI, where two and three binary subsystems respectively are bound, we have for the lowest solution that $E < 0$ and $E_x(\rho \to \infty) < 0$. The angular wavefunction in eq. (25) is exponentially decaying as in the previous case of one bound subsystem. However, now several decoupled (at large $\rho$) solutions exist each corresponding to the bound states of different binary subsystems. The lowest $\tilde{\lambda}$ values diverge as $\tilde{\lambda}(\rho \to \infty) \to -|E_x|\rho^2 2m/\hbar^2$, where the possibly different $E_x$ values correspond to the eigenvalues in the bound subsystems. The higher lying $\tilde{\lambda}$ values again converge to the hyperspherical spectrum for $\rho \to \infty$.

The total wavefunction is like in region IV then asymptotically decaying as a product of two independent subsystems corresponding to the appropriate Jacobi coordinates $x$ and $y$. However, now several different solutions exist each related to the different two-body bound state. Moving accross the lines separating region V from the regions IV and VI all wavefunctions are smooth and consequently also the radial moments are smooth functions of the binding energy.

An interesting anomaly, the Efimov effect, is present when at least two of the potentials simultaneously have bound states at zero energy or equivalently infinite scattering lengths. Then one $\lambda(\rho \to \infty)$ approaches a negative constant, which in turn leads to infinitely many bound state solutions to eq. (8), see ref. [9]. As indicated on fig. 1 these regions appear as series of infinitely long rods in the three dimensional space spanned by the strengths of the short-range potentials. This structure will be discussed in detail in a forthcoming publication [23].

When a long-range two-body potential is present, the asymptotic behavior is characterized exactly as in the case, where one binary subsystem is bound. The Efimov effect is prohibited by the presence of such a potential.

# 5 Numerical results

We shall in this section illustrate the actual numerical behavior of three-body systems in various regions of fig. 1. The general analytical conclusions in the previous subsections were explained by use of square well potentials, but in the numerical illustrations we shall assume Gaussians, $-S \exp(-(r/b)^2)$, for all the short-range two-body potentials. This may be viewed as a slight improvement, which at the same time will demonstrate the generality of our arguments. We shall confine ourselves to two nucleons outside a heavier core.

The potential parameters basically only enter in specific combinations. The range parameters can for example be compensated by adjusting the strength parameters and we shall therefore choose $b_{nn} = 1.8$ fm and the other two identical ranges $b_{nc} = 2.55$ fm as appropriate respectively for the nucleon-nucleon and the nucleon-core potentials, see ref. [13, 24]. The related strengths are left to vary the partial energies of the (sub)system(s). The Coulomb potentials are obtained by assuming point charges for the nucleons and a gaussian charge distribution for the core. The potentials are then respectively proportional to $e^2/r$ and $e^2/r$ erf($r/b_{nc}$). This suffice for our purpose, but needs perhaps adjustments in future accurate applications.

## 5.1 The effective radial potential

The crucial quantity in our formulation is clearly the angular eigenvalue spectrum or the effective radial potentials. Their behavior as function of $\rho$ for the different angular momentum quantum numbers determine the structure of the three-body system at both small and large distances. The characteristic properties of the eigenvalue spectrum is shown in fig. 2a for two neutrons outside a core of mass 9. The hyperspherical spectrum $K(K+4)$ arises at $\rho = 0$ as well as at $\rho = \infty$. The neutron-core potential has, unlike the neutron-neutron subsystem, a bound state at $-0.72$ MeV, which results in the parabolic divergence of the lowest eigenvalues. They correspond asymptotically to the bound binary subsystem with the third particle far away in orbits of angular momentum respectively zero and one.

The imposed symmetry of the spatial wavefunctions of the two neutrons remove two other asymptotically identical and orbitally asymmetric states. The divergences leave room for the lowest hyperspherical levels at $\rho = \infty$. They are replaced by other levels originating from higher lying levels at $\rho = 0$.

The signature of a binary bound state is further illustrated in fig. 2b, where the attractive neutron-neutron potential artificially is made strong enough to hold a bound state at $-2.1$ MeV. This produces additional two parabolically (faster) diverging levels corresponding to the bound binary subsystem and the core far away in orbits of angular momentum zero and one. The replacement of all these diverging levels must now originate from even higher lying levels at $\rho = 0$. The slopes of the levels are also seen to be correspondingly larger.

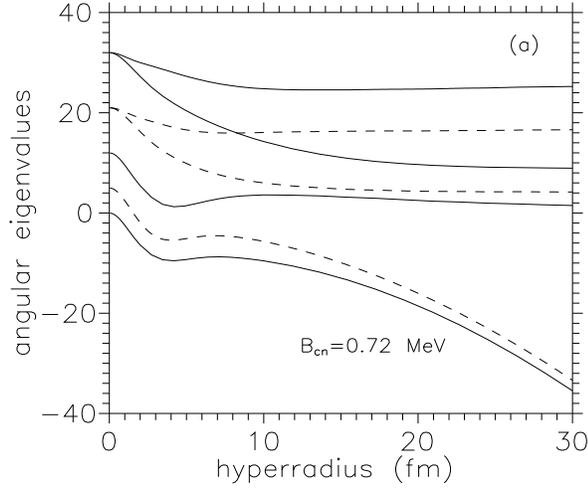
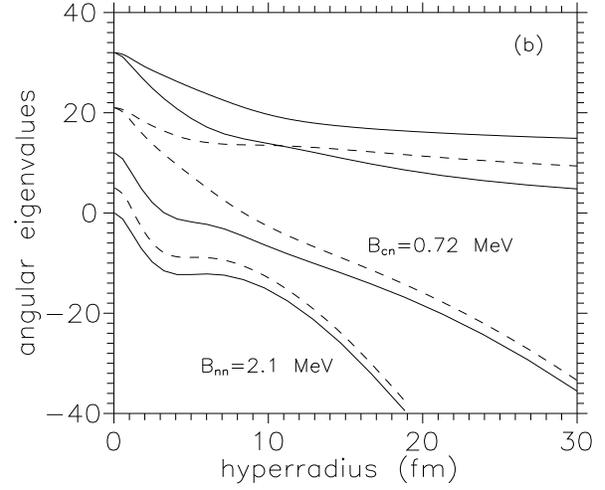

Figure 2a. The angular eigenvalue spectrum $\lambda$ as function of $\rho/\rho_0$ for the nucleon-nucleon and nucleon-core range and strength parameters respectively given by $b_{nn} = 1.8$ fm, $S_{nn} = 31$ MeV and $b_{nc} = 2.55$ fm, $S_{nc} = 15$ MeV. The conserved total (orbital) angular momentum is 0 (solid curves) and 1 (dashed curves). These potentials reproduce the free neutron-neutron low-energy scattering data and correspond to a neutron-core bound state at an energy of $B_{nc} = 0.72$ MeV. Only orbitally symmetric states in nucleon-nucleon interchange is exhibited. Thus a total spin zero is implicitly assumed.

Figure 2b. The same as fig. 2a with $S_{nn} = 52$ MeV corresponding to a nucleon-nucleon bound state at an energy of $B_{nn} = 2.1$ MeV.

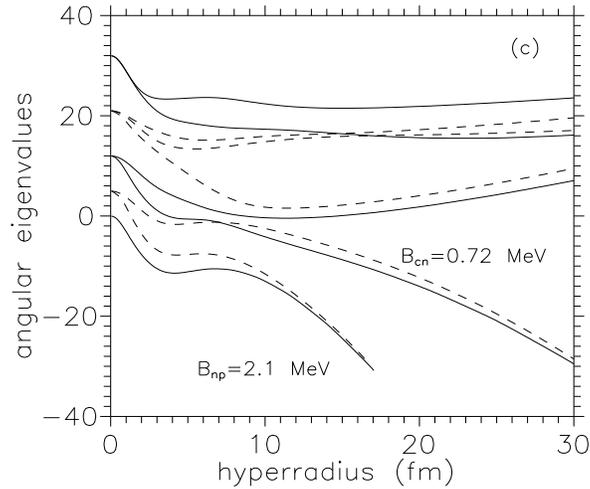

Figure 2c. The same as fig. 2a when the charge of the core and one of the nucleons are respectively 3 and 1 unit of charge, which makes one of the nucleon-core systems unbound and the neutron-proton bound with the energy $B_{np} = 2.1$ MeV. Both orbitally symmetric and antisymmetric states are now exhibited.

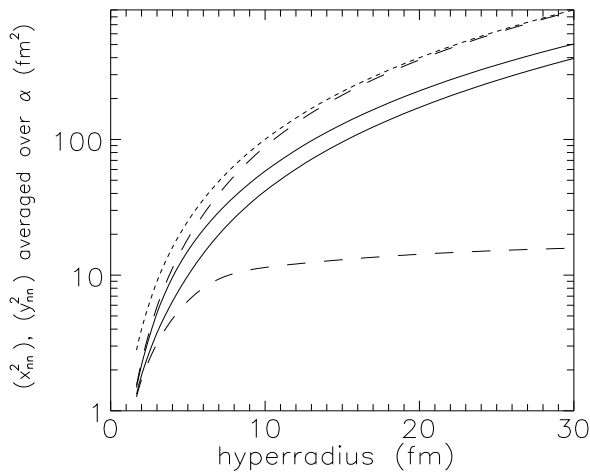

Figure 3. The "angular" expectation values of $x_{nn}^2$ and $y_{(nn)c}^2$ for an angular momentum zero state as function of hyperradius $\rho$ for $S_{nc} = 7.8$ MeV (no bound state) and $S_{nn} = 31$ MeV (no bound state, solid curves where the lowest corresponds to $x^2$) and $S_{nn} = 52$ MeV (one bound state of binding energy 2.1 MeV, dashed curves where the lowest corresponds to $x^2$) corresponding to regions II and IV in fig. 1b. The upper curve (short dashed) is $\rho^2 = x^2 + y^2$.

Inclusion of the Coulomb potential qualitatively changes the eigenvalue spectrum. An example is shown in fig. 2c, where a deuteron-like neutron-proton system interacts with a core by means of a short-range nucleon-core potential with a bound state at $-0.72$ MeV, and a Coulomb potential which makes the proton-core system unbound. More levels now appear due to the absence of symmetry constraints on the wavefunctions. However, asymptotically it is still the same four parabolically diverging eigenvalues corresponding to the binary bound subsystem. The positive eigenvalues diverge as expected linearly towards $+\infty$.

## 5.2 Asymptotic behavior

The radial moments characterize the spatial extension of a system. For the three-body problem the two lengths corresponding to $x$ and $y$ are in principle both needed for a complete description. In the numerical examples we shall only discuss the moments of second order. The natural unit of length when Jacobi coordinates are used is the hyperradius $\rho$ and we shall illustrate the structure by the moments of $\rho$. The substructure is conveniently illustrated by the "angular" expectation values

$$< x_i^2 > = \int d\Omega |\Psi|^2 x_i^2, \quad < y_i^2 > = \int d\Omega |\Psi|^2 y_i^2 ,  \qquad (31)$$

which are functions of $\rho$.

The characteristic structure of a zero angular momentum state can be seen in fig. 3, where these expectation values are shown for two different sets of potentials. When no binary subsystem is bound the radial moments of all Jacobi coordinates diverge proportional to $\rho^2$. For a given hyperspherical quantum number $K$ the ratio is then constant, e.g. equal to unity for $K = 0$. When one subsystem is bound, the corresponding $< x^2 >$ converges towards the finite value for the binary bound subsystem. This reflects the fact that the long-distance asymptotic behavior is decribed by two independent exponentially decaying wavefunctions in the $x$ and $y$ degrees of freedom. The remaining, diverging part of $\rho^2$ is correspondingly taken up by $< y^2 >$.

The structure of the angular expectation values for finite angular momenta are now also easily understood. When no binary subsystem is bound all radial moments of any set of Jacobi coordinates are similar, i.e. diverge or converge in roughly constant ratios as discussed in details in ref. [6]. When one binary subsystem is bound in a state of zero angular momentum, the large-distance behavior is characterized as a two-body system where the third particle is carrying the total angular momentum of the system in its orbital motion. Thus the angular expectation values are asymptotically identical to those of fig. 3.

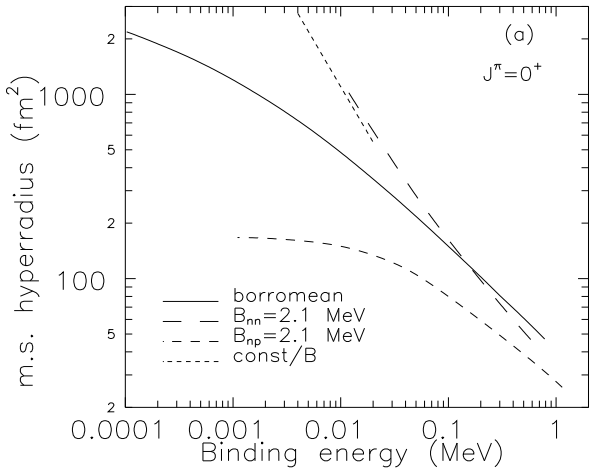 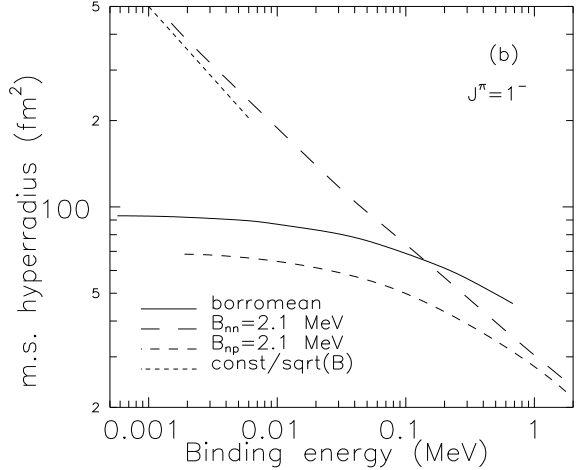

Figure 4a. The mean square hyperradius $<\rho^2>$ for a zero angular momentum state as function of binding energy for two nucleons outside a core of mass 9 and charge 3. The parameters $S_{nn} = 31$ MeV (no bound state, solid curve), $S_{nn} = 52$ MeV (one bound state of binding energy 2.1 MeV, long-dashed curve compared with the asymptotic two-body behavior inversely proportional to the binding energy dotted curve) and $S_{np} = 52$ MeV (one bound state of binding energy 2.1 MeV, short-dashed curve) corresponding to regions II and IV in fig. 1b. The curves are found by varying $S_{nc}$, the short-range interaction between the nucleons and the core.

Figure 4b. The same as fig. 4a for an angular momentum 1 state.

The mean square hyperradius for a zero angular momentum state is shown in fig. 4a as function of binding energy for different systems. The logarithmic divergence is seen for the borromean system (no bound subsystems) whereas the two-body asymptotic behavior arises when one or more of the subsystems are bound. The inverse proportionality with the binding energy when only short-range interactions are present and the finite value due to the Coulomb confinement are both reached at about 10 keV. Both asymptotics are needed for binding energies above 10-50 keV when a binary subsystem is bound by 2 MeV. The influence of angular momentum is shown in fig. 4b for different cases of angular momentum one states. Both three-body and two-body asymptotics are seen with the respective converging and diverging behavior.

It is important to realize that the concept of halo structures are not restricted to the ground states of the systems. Excited states exhibiting halo properties are probably much more abundant. The asymptotic behavior of such states are fortunately easily explained, since the low energy dependence is identical to that of the ground state. Thus, when the normalization as well as the radial moment in question diverge, the normalized radial moment is determined entirely by tail properties and is therefore identical for ground- and excited states. When conversely the normalization converges and consequently is determined at smaller distances, the size of the tail, but not the decay rate, might differ for the two states. Thus, any radial moment has the same functional energy dependence, but the proportionality constants are different.

The vanishing binding energy for the excited state can be achieved by varying either $S_{nn}$ or $S_{nc}$, i.e. respectively horizontal or vertical variation on fig. 1b. The arguments above are valid for states outside the shaded region in this figure. The approach of the Efimov region (varying $S_{nc}$) brings in a very different type of asymptotic behavior [9, 23].

# 6  Summary and conclusions

We have recently investigated the asymptotic properties of nuclear three-body bound states arising through weak binary interactions, which are unable to bind the individual subsystems ref. [6]. These investigations are extended in the present paper to include three-body systems, where one or more of the binary subsystems are bound. The major complication is due to the mixture of two- and three-body asymptotic large-distance behavior.

We first generalized the previous method of the kinetic hyperspherical expansion, where the complete basis set is the eigenfunctions of the kinetic energy operator in Jacobi coordinates. The basis set is now the angular eigenfunctions of kinetic energy plus (at least) the strongest of the two-body interactions. The proper asymptotics is then obtained in this Schrödinger formulation provided the two weakest two-body potentials are unable to bind their respective subsystems.

When more than one of the binary potentials have bound states, Faddeev equations are needed and we formulate and exploit a new method, where the angular part of the wavefunctions first is computed for each of the three components. The set of coupled radial equations is the same in the Schrödinger and the Faddeev formulations. All possible asymptotics, including the Efimov anomaly, can now be described and the many different cases are then catalogued and discussed.

When none of the subsystems have bound states, we recover all the results previously obtained, i.e the mean square radius is at most logarithmically divergent. When exactly one subsystem have a bound state of energy $E_2$ and the total three-body energy approaches $E_2$, the system behaves as one particle in the common field of the two-body state, i.e. the asymptotics and the related possible divergences are those of a two-body system. Radially excited states above the two-body threshold $E_2$ are sometimes possible and their asymptotics are of genuine three-body nature, as if the bound subsystem did not exist. Angular momentum and Coulomb barriers influence such systems in accordance with the two- or three-body nature of the states as described in previous publications ref. [4, 6].

When at least two subsystems have bound states, the halo structure is not possible in the ground state, which necessarily is too strongly bound and consequently spatially confined. The most spectacular structure arises in excited states, when at least two of these bound state energies simultaneously approaches zero. These are the spatially extended and weakly bound Efimov states. They are strongly hindered by finite angular momenta and prohibited by Coulomb potentials. Apart from the Efimov singularity other more ordinary halo structure may arise in excited states, where the three-body energy approaches zero and the structure is of genuine three-body character.

The actual rate of divergence and the related nuclear structures are numerically investigated in detail for two nucleons outside a core. When one binary subsystem is bound by 2 MeV the two-body asymptotics is reached at about 10 keV. A mixture of two- and three-body asymptotics are necessary at somewhat larger energies. Less binding in the subsystem decreases the three-body binding energy below which only two-body asymptotics is needed.

When two nucleons outside a core is the ground state of a nucleus and furthermore is weakly bound, we are on the driplines. Here the residual interaction plays an important role and the "core plus one nucleon" systems are almost always unbound, i.e. we are in region II in fig. 1. The exceptions — $^9$C and $^{13}$O on the proton dripline, $^{23}$N and $^{24}$O on the neutron dripline — are typically bound by several MeV and interesting effects are more likely to occur in excited states. For excited states we would in general expect the good cases to be just below the two-nucleon threshold, but above the one-nucleon threshold, i.e. in region V in fig. 1b. Close to stability such states, however, lie in an energy range where the level density is high and the states are therefore more easily mixed with other states. Further towards the dripline such states have not yet been reached in experiments.

In conclusion, weakly bound nuclear three-body systems may exhibit halo structure either in ground states or in excited states. Various types of both two-body and three-body character are possible depending on the bound states of the two-body subsystems. This investigation has attempted a classification of the possible nuclear halo structures.

**Acknowledgments** One of us (DVF) acknowledges the support from the Danish Research Council.

# Appendix. Jacobi- hyperspherical- and hyperangular coordinates

We consider a system of three particles with masses $m_i$ and coordinates $\mathbf{r}_i$. The Jacobi coordinates are defined as

$$\mathbf{x}_i = a_{jk}\mathbf{r}_{jk}, \qquad \mathbf{y}_i = a_{(jk)i}\mathbf{r}_{(jk)i}$$
$$a_{jk} = \left(\frac{1}{m}\frac{m_j m_k}{m_j + m_k}\right)^{1/2}, \qquad a_{(jk)i} = \left(\frac{1}{m}\frac{(m_j + m_k)m_i}{m_1 + m_2 + m_3}\right)^{1/2} \qquad (1)$$
$$\mathbf{r}_{jk} = \mathbf{r}_j - \mathbf{r}_k, \qquad \mathbf{r}_{(jk)i} = \frac{m_j\mathbf{r}_j + m_k\mathbf{r}_k}{m_j + m_k} - \mathbf{r}_i \; .$$

where $\{i, j, k\}$ is a cyclic permutation of $\{1, 2, 3\}$ and $a_{jk}^2$ is the reduced mass of the $j$'th and $k$'th subsystems in units of $m$, which is a normalization mass chosen to be equal to the nucleon mass.

The hyperspherical variables [25] are introduced as

$$\rho, \ \mathbf{n}_{x_i} = \mathbf{x}_i/|\mathbf{x}_i|, \ \mathbf{n}_{y_i} = \mathbf{y}_i/|\mathbf{y}_i|, \ \alpha_i, \tag{2}$$

where $\alpha$, in the interval $[0, \pi/2]$, is defined by

$$\rho^2 = \mathbf{x_i}^2 + \mathbf{y_i}^2, \ |\mathbf{x_i}| = \rho \cos \alpha_i, \ |\mathbf{y_i}| = \rho \sin \alpha_i . \tag{3}$$

We omit the indices where we need not emphasize the particular set of Jacobi coordinates. Note that $\rho$ is independent of what set is used.

The hyperangular coordinates are now defined by $\{\rho, \Omega_i\} = \{\rho, \alpha_i, \Omega_{ix}, \Omega_{iy}\}$, where $\Omega_{ix}$ and $\Omega_{iy}$ are the angles describing the directions of $\mathbf{x_i}$ and $\mathbf{y_i}$.

The relation between three different sets of Jacobi coordinates is given by

$$\mathbf{x}_k = \mathbf{x}_i \cos \phi_{ik} + \mathbf{y}_i \sin \phi_{ik}, \ \mathbf{y}_k = -\mathbf{x}_i \sin \phi_{ik} + \mathbf{y}_i \cos \phi_{ik} \tag{4}$$

where the transformation angle $\phi_{ik}$ is given by the masses as

$$\phi_{ik} = \arctan \left( (-1)^p \sqrt{\frac{m_j (m_1 + m_2 + m_3)}{m_k m_i}} \right) \tag{5}$$

where $(-1)^p$ is the parity of the permutation $\{i, k, j\}$.